# Guáman Poma's Yupana and Inca Astronomy

Subhash Kak
Oklahoma State University, Stillwater, OK 74078, USA

**Abstract.** An explanation is provided for the Inca counting board described by Guáman Poma in 1615. Although the board could have been used in more than one way, we show that based on certain reasonable assumptions regarding non-uniform representation of numbers its most likely use was counting in multiples of 6, 24, and 72. The independent numbers represented on its five rows are 92, 31, 29, 79, and 56 that appear to be astronomically connected to sub-periods within the year and planet periods in a manner similar to Mayan astronomy. Based on these and other considerations we propose that the board fulfilled an astronomical counting function.

I.  INTRODUCTION

This article explains the logic of the Inca calculating board (*yupana,* Figure 1) shown in the book *El Primer Nueva Corónica y Buen Gobierno* (The First New Chronicle and Good Government), a Peruvian chronicle completed around 1615 by Felipe Guáman Poma de Ayala, and sent as a handwritten manuscript to King Philip III of Spain (Guáman Poma, 1615). The counting use of the grid of solid and unfilled dots or tokens appears reasonable because of its appearance with a khipu (or *quipu*) (Figure 2) that was widely used in the Inca world for recording numbers (and other information) (Urton, 2003). That the *yupana* (Quechua for the verb to count) is not a simple abacus that does decimal computations is clear from the constraints that are placed on the number of pieces or tokens in each position.

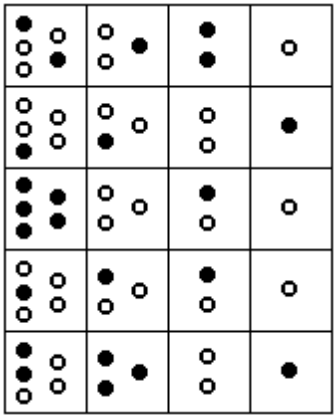
Figure 1. Yupana of Guáman Poma

The understanding of Guáman Poma's Yupana (henceforth GPY) is of considerable importance in making sense of Inca astronomy. Several theories have been proposed for the use of this yupana but none of them is adequate. Some of these theories take the numbers in the higher row to be 10 times the one in the immediately lower row (Wassen, 1931; Radicati de Primaglio, 1979; Burns Glynn, 1981) with different weights for the four columns.

Wassen takes the weights to be 1, 5, 15, and 30 from left to right which means that the total sum of the lowest row is 80. This is an arbitrary procedure leading to no advantage in calculations. The theories of Radicati and Burns—who takes the rightmost column to be memory—take the weights to be the same but they don't have explanation for the use of columns. This also does not explain why each row has 11 pieces and why tokens of the same value are lumped together in groups of 5, 3, 2, and 1. Although GPY could have been used as an abacus (Leonard and Chakiban, 2010), the fact of its four divisions for each digit argues against that being its primary function.

Two other arbitrary approaches take the columns to be in the proportion 1, 2, 3, 5 (Mendizabal, 1971), and 5, 3, 2, 1 (De Pasquale, 2001). Like the other previous proposals, these do not provide unique mapping of numbers to token placements which rules them out as valid solutions. Yet another theory (Florio, 2009) takes the rightmost column to be multiplicand and the next two columns to be multipliers and the leftmost column to be the sum of the second and the third columns. This theory further takes the dark circle to be 1 and the light circle to be 10. This representation is arbitrary without any specific advantage in computations and, furthermore, it doesn't even check out for the dots on Poma de Ayala's yupana. Florio explains that by claiming that Poma de Ayala's figure is wrong.

It is possible that GPY is a device that had multiple uses based on the mathematical expertise of the user just as calculating devices like the slide rule can be used in different ways. But what we are interested in is its function that is matched to the four columns on the board. All the previous proposals fail this test and they are not intuitively reasonable.

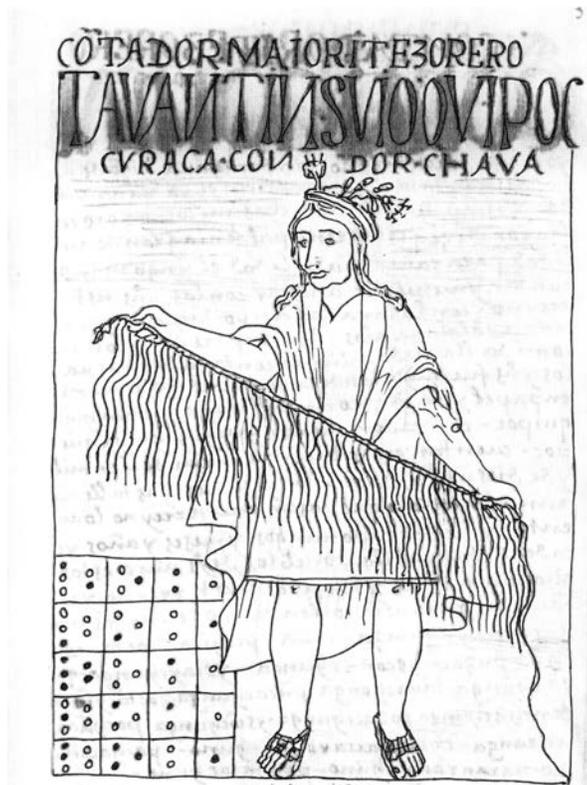
Figure 2. Guáman Poma's illustration of yupana and khipu

The theories that consider a positional system in powers of 10 are motivated by the knowledge that in Quechua language (which was the language of the Inca Empire) number words use base 10. Here is a short list of number words from Huallaga Quechua (Gildorf, 2000):

  *ch'usaq* – 0, *huk* - 1, *iskay* - 2, *kimsa* - 3, *tawa* - 4, *pichqa* - 5, *suqta* - 6, *qanchis* - 7, *pusaq* - 8, *isqun* - 9;

    *chunka* - 10, *pachak* - 100, *waranqa* - 1000.

For more complex number words, the nucleus is always a power of 10. Thus *isqon pachak* is 900; *qanchis chunka pichqa* is 7; and *kimsa pachak tawa chunka qanchis waranqa iskay* is 347,002.

We have evidence that the Inca were very proficient at mathematics, as from the claim of the Spanish priest José de Acosta who lived amongst them and wrote *Historia Natural Moral de las Indias*:

> To see them use another kind of quipu, with maize kernels, is a perfect joy. In order to carry out a very difficult computation for which an able computer would require pen and paper, these Indians make use of their kernels. They place one here, three somewhere else and eight, I know not where. They move one kernel here and there and the fact is that they are able to complete their computation without making the smallest mistake. as a matter of fact, they are better at practical arithmetic than we are with pen and ink. Whether this is not ingenious and whether these people are wild animals let those judge who will! What I consider as certain is that in what they undertake to do they are superior to us. (Acosta, 1596; quoted also in Joseph, 2000)

Numerical information associated with administrative, military, and business uses was stored in the knots of khipu (quipu) strings in base-10 digits as in the Quechua language. Numerical calculations were done on a variety of yupanas in which piles of tokens, seeds, or pebbles were moved between different compartments of the yupana for the calculation to take place. The Inca also performed cyclic computations and multiplication with fractions (Ascher, 1983).

We expect, therefore, that Pomo's yupana was based on sound mathematical principles – in addition to being intuitive–and this paper presents a new explanation based on non-uniform representation of numbers. We show that this mathematical basis yields numbers that are plausibly connected to Inca astronomy.

## II. NON-UNIFORM REPRESENTATION OF NUMBERS

In our familiar mathematics with Indian numerals (also in khipu), each place can have one of 10 different numbers. This constitutes a uniform representation system in which the total count as one increases the size of the system to, say, v places is simply $10^v - 1$. In v = 4 as below, then number of tokens in each cell is 10.

| 10 | 10 | 10 | 10 |
|---|---|---|---|

Figure 3. Base-10 representation

The largest four-digit number than can be counted is $10^4 - 1 = 9999$. The weights associated with the 10 different digits are shown in the row below, where the least significant digit is to the right:

| | 10 | 10 | 10 | 10 |
|---|---|---|---|---|
| Weights: | $10^3$ | $10^2$ | 10 | 1 |

Figure 4. Base-10 representation with weights

The GPY, however, is a non-uniform representation system since the squares in each row have different number of tokens (Figure 5). The counting scheme must take this fact into consideration.

| 5 | 3 | 2 | 1 |

Figure 5. Number of tokens in each row of GPY

We take the count to be zero for a token that is not fully dark (implying it has not been placed in the square) and one if it is. Note also that it is intuitive to take the left-most square as the least significant place because of the five tokens assigned there. Incidentally, this is how the assignment is done in the Chinese abacus *suànpán* as well.

To understand the workings of a non-uniform number representation system, consider now the 6-place system given below.

| Columns: | I | II | III | IV | V | VI |
|---|---|---|---|---|---|---|
|  | j | k | m | n | p | q |

Figure 6. A 6-place, non-uniform representation system

For it to count numbers in a regular fashion, the weights will have to be as follows:

Column I: The weight should be 1, and the total count achieved by it is $j$.

Column II: The weight should be (j+1) since numbers up to j have been counted by the previous column. The total count achieved by Columns I and II is $(j+1)k + j = (j+1)(k+1) - 1$.

Column III: This column should count starting with 1 more than the total count of Columns I and II. The weight of each token here is, therefore, $(j+1)(k+1)$. The total count in Columns I, II, and III is $(j+1)(k+1)m + (j+1)(k+1) - 1$, which is $(j+1)(k+1)(m+1) - 1$.

We can now generalize the count and conclude that the total count in Column VI is: $(j+1)(k+1)(m+1)(n+1)(p+1)(q+1) - 1$. This constitutes the proof of the theorem below.

**Theorem.** For a non-uniform number representation system with *s* places, with the number of tokens in the positions counting from the least significant to be $n_1, n_2, ... n_s$, the total count is $\prod_{i=1}^{s}(n_i + 1) - 1$.

### III. ANALYSIS OF GPY

The Inca society had expert record keepers known as *khipukamayuq* whose work involved much abstraction. To the main cord were attached pendant strings some of which were attached to secondary and tertiary strings (Boone and Urton, 2011). The number of pendants can be as high as 1,500. In a recently discovered collection of 32 khipu, the range of knots varies from 6 to 762, with the average being 149 (Urton, 2003).

The knots on the strings were tied in patterned arrangements that indicated higher powers of ten. The knots also came in different colors. Clearly, the record-keepers in the Inca society dealt with complex numerical and other information. Given this, we consider the following assumptions to determine the workings of GPY to be reasonable:

- It represents numbers in a unique fashion
- Systematic mathematical operations can be made on it
- It represents numbers efficiently

Note that the left two columns of GPY have elements in two columnar groups. It is possible for the items to have had different positional value in each of these sub-columns, but we begin by assuming that the items were lumped together and represented in two columns only for aesthetic purposes.

## IV. BASIC FUNCTION GPY

In the basic function GPY, the board has 4 columns. The lowest row of GPY is described by j, k, m, and n (or $n_1, n_2, n_3, n_4$ values of Theorem) of 5, 3, 2, and 1, the total count in this row is $(5+1)(3+1)(2+1)(1+1) - 1 = 144 - 1 = 143$.

| | 5 | 3 | 2 | 1 |
|---|---|---|---|---|
| Weight: | 1 | 6 | 24 | 72 |

Figure 7. Basic GPY counting in the lowest row

| Number | Yupana | Columns | | |
|---|---|---|---|---|
| 1 | 1 | 0 | 0 | 0 |
| 2 | 2 | 0 | 0 | 0 |
| 3 | 3 | 0 | 0 | 0 |
| 4 | 4 | 0 | 0 | 0 |
| 5 | 5 | 0 | 0 | 0 |
| 6 | 0 | 1 | 0 | 0 |
| 7 | 1 | 1 | 0 | 0 |
| . | . | . | . | . |
| 18 | 0 | 3 | 0 | 0 |
| 19 | 1 | 3 | 0 | 0 |
| . | . | . | . | . |
| 23 | 5 | 3 | 0 | 0 |
| 24 | 0 | 0 | 1 | 0 |
| . | . | . | . | . |
| 143 | 5 | 3 | 2 | 1 |
| Weights: | 1 | 6 | 24 | 72 |

Figure 8. Basic GPY counting up to 143

This counting may be continued in the higher rows, where the computation is in powers that are increased by $12^2 = 144$ as one climbs each row. This would be the case if the board is used to represent a single number. For such a case, the board together with values for each row and that of the tokens is shown in Figure 9.

| | | | | |
|---|---|---|---|---|
| $12^8$ | 2 | 1 | 2 | 0 |
| $12^6$ | 1 | 1 | 0 | 1 |
| $12^4$ | 5 | 0 | 1 | 0 |
| $12^2$ | 1 | 1 | 0 | 1 |
| $12^0$ | 2 | 3 | 0 | 1 |
| Weights: | 1 | 6 | 24 | 72 |

Figure 9. Basic GPY token values for each row
if used to represent a single number

It is significant that counting by 6, 24 and 72 is basic to GPY. We do know that the Incas used 12 solar and lunar months.

| | | | | |
|---|---|---|---|---|
| $12^2$ | 5 | 0 | 0 | 0 |
| $12^0$ | 0 | 0 | 0 | 0 |

Figure 10. The number 720

Mathematical operations can be done easily on the yupana. Let us wish to add 57 and 49. They would be represented as in Figure 11 in the bottom two rows with the sum in the top row:

| | | | | |
|---|---|---|---|---|
| 4 | 1 | 1 | 1 | 106 |
| 3 | 1 | 2 | 0 | 57 |
| 1 | 0 | 2 | 0 | 49 |

Figure 11. Adding 49 and 57

The addition operation simply involves adding the corresponding numbers in each column and once the maximum of a specific column has been exceeded, taking the extra digit to the next column on the right. Thus in the example above the third column maximum is 2 and since the column to its right equals three tokens from it, the total of 2+2 = 4 is distributed as a carry of 1 (which equals 3) and the remainder of 1, which remains in the column.

Similarly, subtraction can be done in 1s, 6s, 24s, and 72s. Performing addition and subtraction repeatedly implies the capability to do multiplication and division.

V. THE GPY NUMBERS

In the general case, the numbers in the different rows could all be in the same basis and used, for example, for representation of independent numbers. If that was done, then the tokens on GPY correspond to the numbers shown in Figure 12.

| 2 | 1 | 2 | 0 | 56 |
|---|---|---|---|----|
| 1 | 1 | 0 | 1 | 79 |
| 5 | 0 | 1 | 0 | 29 |
| 1 | 1 | 1 | 0 | 31 |
| 2 | 3 | 0 | 1 | 92 |

| Weights: | 1 | 6 | 24 | 72 |
|---|---|---|---|---|

Figure 12. Numbers on GPY

The five numbers add up to 287, which is 78 short of 365, the duration of the solar year.

The number 78 is important in ancient astronomy for being exactly 1/10 of the synodic period of Mars. If we take that as a clue then we would like to determine if the five numbers mentioned in Figure 12 also have connections with astronomical facts likely to have been known to the Incas. But before we do so we examine whether the GPY was used in an advanced 6-column mode.

## VI. POSSIBLE ADVANCED USAGE

In the advanced function GPY, the board has 6 columns. The lowest row of the advanced function GPY is described by j, k, m, n, p, and q (or $n_1, n_2, n_3, n_4, n_5$ values of our Theorem) of 3, 2, 2, 1, 2, and 1, the total count in this row will be $(3+1)(2+1)(2+1)(1+1)(2+1)(1+1) - 1 = 432 - 1 = 431$.

| | 3 | 2 | 2 | 1 | 2 | 1 |
|---|---|---|---|---|---|---|
| Weights: | 1 | 4 | 12 | 36 | 72 | 216 |

Figure 13. Advanced GPY token values for each row

The number 6 years equal to 2,160 days is represented in this mode as in Figure 14:

| | 5 | 0 | 0 | 0 | 0 | 0 |
|---|---|---|---|---|---|---|
| | 0 | 0 | 0 | 0 | 0 | 0 |
| Weights: | 1 | 4 | 12 | 36 | 72 | 216 |

Figure 14. The number 2,160 (6 years of 360 days each)

When the GPY row numbers are taken to be independent in this mode, they turn out to be 278, 85, 83, 229, and 185 with the total of 860. These numbers appears random without any specific significance and, therefore, we believe that GPY was not used in the 6-column mode.

VII. ASTRONOMICAL CONTENT OF GPY NUMBERS

Inca astronomers computed equinoxes, solstices, and likely zenith passages. Their calendar was lunisolar for they kept parallel counts for lunar and solar months. As twelve lunar months are 11 days shorter than the full 365-day solar year, those an adjustment was needed to mark the winter solstice. A system of solar horizon markers was used to time equinoxes and solstices. These markers were stone structures of considerable size (Dearborn, 2000). The names of the months are provided in Guáman Poma's book. There is other evidence related to the calendar that includes the ceque system and reports in other early Spanish texts and on khipus (Bauer, 1998; Zuidema, 2005).

Details of Inca astronomy are also provided by the early 17$^{th}$ century secret Jesuit manuscript *Exsul Immeritus Blas Valera Populo Suoi* that recently came to light (Laurencich-Minelli and Magli, 2009/2010; Zuidema, 2005). Figure 15 summarizes the counts of the months of the Inca year represented by the use of knots on 13 pendants connected to the main cord as described in the *Pachakhipu,* a sheet of paper inserted in *Exsul Immeritus.*

| Month | 1 | 2 | 3 | 4 | 5 | 6 | 7 | 8 | 9 | 10 | 11 | 12 | 13 |
|---|---|---|---|---|---|---|---|---|---|---|---|---|---|
| Days | 29 | 30 | 29 | 29 | 30 | 29 | 30 | 30 | 30 | 29 | 30 | 30 | 10 |

Figure 15. The twelve Inca months according to Pachakhipu; the 13$^{th}$ column is for intercalary days.

In additional to the monthly count, the days were counted in groups of 15 each by the use of alternate colors of red (*hanan*) and green (*hurin*) knots and in groups of 10 by spacing the knots. In the last pendant, the 13$^{th}$, there are five green knots, which represent the days needed to bring the total to 360 (due to the presence of 5 synodic months of 29 days) and five red further knots representing additional intercalary days, needed to bring the solar count to 365. I would like to propose that the extra 10 intercalary days were seen as harmonizing the 354 days of the lunar year (approximated by 355 so as to leave room for the extra *week* of 10 days) with the 365 solar days.

We come back to the use of GPY in its four-column standard mode and examine the five numbers of GPY, namely 92, 31, 29, 79, and 56, going from bottom to top. In a previous section, it was suggested that the number 78 (implied by its absence) is one-tenth the Mars period. Other numbers should then be related similarly to other planet periods. If the Mars number of 78 is a primary number, then only four of the five GPY numbers should be independently significant, since the fifth would be fixed by the constraint of the sum being equal to 365.

Note first that the Inca had strong traditions of counting by both the sun and the moon. Given the clarity of the Andean night, and the Inca mapping of the sky into the sacred landscape of Tawantinsuyu (the Inca Empire), one would expect that the periods of the planets were known and represented in terms of coded relationships of the synodic periods.

If one were to see Inca astronomy as having parallels with Mayan astronomy, one would expect the knowledge of the following numbers:

> 949-day cycle: relating Venus and Sun and equal to the sum of 584-day synodic period of Venus and 365-day period of Sun. This shows up in the long count calendar of 18,980 days (Thompson, 1960).

819-day cycle: related to mean synodic periods of Jupiter and Saturn which was seen in the Maya world as incorporating the Jupiter and Saturn periods of 399 and 378 in the following way:

$$63 \times 6 = 378; \quad 63 \times 19 = 399 \times 3 = 1197 = 819 + 378$$

The 819-day period is in the Dresden codex. The evidence for the 819 cycle is from Palenque (Chiapas) on a stucco panel commemorating an event in the life of the ruler Pacal II in the year 668 (Lounsbury, 1978).

One would also expect to see evidence of the knowledge of the asymmetric quarters of the year due to the very precise determination of the solstice and equinox points. The lengths of the four quarters in the Southern Hemisphere are:

Autumn equinox to Winter solstice: 92 days
Winter solstice to Spring equinox: 93 days
Spring equinox to Summer solstice: 91 days
Summer solstice to Autumn equinox: 89 days

The correct synodic periods of the five planets are:

Mercury: 116; Venus: 584; Mars: 780; Jupiter: 399; and Saturn: 378.

Our proposal for the GPY numbers is as follows:

Number 78. *Mars*, whose synodic period is 780. This is the sixth number associated with GPY by virtue of the constraint of the sum being equal to 365.

Number 56, at the top of the yupana. *Jupiter* and *Saturn*. Jointly through the number 819 as in the Maya system through the equation:

$$56 \times 117 = 819 \times 8$$

This interpretation rests on special significance being given to 8 cycles of the 819-day period. Alternatively, the Inca may have held to the theory that 56 codes the synodic periods of Jupiter and Saturn because its factors 8 and 7 almost exactly divide 399 (Jupiter) and 378 (Saturn), respectively.

Number 79. *Venus*. The synodic period of Venus is 584. The Maya represented this information by considering $584 + 365 = 949$. The Inca appear to have done so by $79 \times 12$. That there is a discrepancy of 1 in this product is a weakness of our interpretation but there may have been an astronomical reason in taking the sum to be 948.

Number 29. *Mercury*: Synodic period is $29 \times 4 = 116$.

Number 31. This number is astronomically significant in the sense of being the length of certain months as in the ceque system (Zuidema, 2005). But this should be seen only as a coincidence and, in our view, this was not one of the primary numbers.

Number 92. *Quarter of the year* from the Autumn equinox to Winter solstice.

This above interpretation of the five GPY numbers is being made as no more than a preliminary hypothesis that needs further elaboration and confirmation. The interpretation as it stands has the weakness of the approximation of 1 in the products associated with Jupiter and Venus. On the other hand, the Inca made other adjustments of one day for the sake of symmetry related to knowledge they indubitably possessed. I refer here to the apparent count of 355 for the lunar year (rather than the more accurate 354) in the khipu calendar where the intercalary month on the 13$^{th}$ pendant has 10 days (Figure 15).

VIII. CONCLUSIONS

This paper has proposed that the yupana of Guáman Poma most likely served the purpose of counting days for calendric use. The unique number representation associated with the yupana is non-uniform that counts in units, sixes, twenty-fours, and seventy-twos. We show that this representation lends itself to intuitive procedures for mathematical operations. A parallel may be drawn with ritual calendrics in India where also a lunisolar system was used and intercalation was required to harmonize the solar and lunar years and different priests were responsible for different counts (Caland, 1982; Kak, 2000).

The non-uniform number representation system associated with GPY leads to the numbers 92, 31, 29, 79, and 56. Based on certain parallels with Mayan astronomy, we have proposed that these numbers imply knowledge of the synodic periods of the planets and the unequal quarters of the year. Even if the Mayan calendar counts were known to the Inca, this system was not adopted since the counting schemes in the two civilizations were different.

If our proposal on the astronomical basis of the five GPY numbers is correct, corroborating evidence should exist in the ceque system and in calendric khipus. There is also the further possibility that the planetary period counts were not derived from the Mayan system and it constituted an independent tradition.

BIBLIOGRAPHY

Acosta, José de. *Historia natural moral de las Indias*. Madrid, 1596. Duke University Press, 2002.

Ascher, Marcia. The logical-numerical system of Inca quipus. *Annals of the History of Computing* 5.3: 268-78, 1983.

Bauer, Brian S. The Sacred Landscape of the Inca: The Cusco Ceque System. Austin: University of Texas Press, 1998.

Boone, Elizabeth Hill and Urton, Gary (eds.). *Their Way of Writing: Scripts, Signs and Pictographies in Pre-Columbian America*. Washington: Dumbarton Oaks Research Library, 2011.

Burns Glynn, William. *La Escritura de los Incas*. Boletín de Lima, No. 12, 13, 14. Lima: Editora Los Pinos, 1981.

Caland, W. (ed. and tr.) *Pañcaviṃśa Brāhmaṇa*. Calcutta: The Asiatic Society, 1982.

Dearborn, David S. P. The Inca: Rulers of the Andes, Children of the Sun. *Astronomy Across Cultures, the History of Non‑Western Astronomy*. Helaine Selin (ed.). Dordrecht: Kluwer Academic Publishers, 197-224, 2000.

De Pasquale, Nicolino. Il volo del condor. *Pescara Informa*, 2001. See also, http://www.quipus.it

Florio, C. *Incontri e Disincontri Nella Individuazione di una Relazione Matematica Nella Yupana in Guáman Poma de Ayala*, Salerno, 14-15 maggio e 10-12 Dicembre 2008 - Oédipus Editore, 2009.

Gilsdorf, Thomas E. Ethnomathematics of the Inkas. *Mathematics Across Cultures, the History of Non‑Western Mathematics*. Helaine Selin (ed.). Dordrecht: Kluwer Academic Publishers, 189-203, 2000.

Guáman Poma de Ayala, Felipe. *El Primer Nueva Corónica y Buen Gobierno (1615)*. Mexico City: Siglo XXI Editores, 1980.

Joseph, George G. *The Crest of the Peacock: Non‑European Roots of Mathematics, New Edition*. Princeton: Princeton University Press, 2000.

Kak, Subhash. Birth and early development of Indian astronomy. *Astronomy Across Cultures, the History of Non‑Western Astronomy*. Helaine Selin (ed.). Dordrecht: Kluwer Academic Publishers, 303-340, 2000.

Laurencich-Minelli, Laura and Magli, Giulio. A calendar Quipu of the early 17[th] century and its relationship with the Inca astronomy. *Archaeoastronomy* 22, 2009/2010.

Leonard, Molly and Chakiban, Cheri. The Incan abacus: A curious counting device. *J. of Math. and Culture* 5(2): 81-106, 2010.

Lounsbury, Floyd G. Maya numeration, computation and calendrical astronomy. *Dictionary of Scientific Biography*. Vol. 15, Supplement 1. New York, 1978.

Mendizabal Losack, Emilio. *Estructura u Funcion en la Cultura Andina. Fase Inka*. Tesos doctoral, Universidad Nacional Mayor de San Marcos, Lima, 1971.

Radicati de Primeglio, Carlos. *El Sistema Contable de los Incas.* Libreria Studium, Lima, 1979.

Thompson, J. Eric S. *Maya Hieroglyphic Writing: An Introduction*. Norman: University of Oklahoma Press, 1960.

Urton, Gary. *Signs of the Inka Khipu: Binary Coding in the Andean Knotted String Records*. Austin: University of Texas Press, 2003.

Wassen, Henry. The Ancient Peruvian Abacus. *Comparative Ethnographical Studies*, vol. 9, E. Nordenskiold (ed.), Goteborg, 1931.

Zuidema, R. Tom. The Inca calendar, the ceque system, and their representation in *Exsul Immeritus*. 75-104, 2005.